\documentclass[aps,twocolumn,footinbib,
letterpaper,showpacs]{revtex4}
\usepackage{natbib}
\usepackage{times}
\usepackage{graphicx}
\usepackage{epsfig}
\usepackage{subfigure}
\usepackage{amssymb,amsmath}
\usepackage{hyperref}
\usepackage{setspace}
\usepackage{url}

\hypersetup{
    colorlinks = true,
    linkcolor = blue,
    anchorcolor = blue,
    citecolor = blue,
    filecolor = blue,
    pagecolor = blue,
    urlcolor = blue
}

\begin{document}

\title{CDMS II Inspired Neutralino Dark Matter in Flipped SU(5)}

\author{Ilia Gogoladze\footnote{On leave of absence from: Andronikashvili Institute of Physics, GAS, \\ Tbilisi, Georgia.}}
\email{ilia@bartol.udel.edu}
\affiliation{Bartol Research Institute, Department of Physics and Astronomy\\
University of Delaware, Newark, Delaware 19716}

\author{Rizwan Khalid}
\email{rizwan@udel.edu}
\affiliation{Bartol Research Institute, Department of Physics and Astronomy\\
University of Delaware, Newark, Delaware 19716}

\author{Shabbar Raza\footnote{On study leave from: Department of Physics, FUUAST, \\ Islamabad, Pakistan.}}
\email{shabbar@udel.edu}
\affiliation{Bartol Research Institute, Department of Physics and Astronomy\\
University of Delaware, Newark, Delaware 19716}

\author{Qaisar Shafi}
\affiliation{Bartol Research Institute, Department of Physics and Astronomy\\
University of Delaware, Newark, Delaware 19716}

\pacs{12.60.Jv, 12.10.Dm, 14.80.Ly}

\begin{abstract}

We investigate neutralino dark matter in supersymmetric flipped SU(5), 
focusing on candidates with masses of order $30$ - $150\,{\rm GeV}$ and
spin independent cross sections that are consistent with the most recent 
CDMS II results. We assume gravity mediated supersymmetry breaking and 
restrict the magnitude of the soft supersymmetry breaking mass 
parameters to 1 TeV or less. With non-universal soft gaugino and Higgs 
masses, and taking flipped SU(5) into account, we  
identify allowed regions of the parameter space and 
highlight some benchmark solutions including Higgs and 
sparticle spectroscopy. 

\end{abstract}

\maketitle


In a recently published paper \cite{cdms:2009zw} the Cryogenic Dark Matter Search 
experiment reported two events in the signal region and a combined upper 
limit, at 90\% confidence level, of $3.8 \times 10^{-44} {\rm cm}^2$
on the spin independent WIMP-nucleon elastic-scattering cross section 
for a WIMP of mass 70 GeV. These events may be interpreted 
as the scattering of the dark matter (DM) from germanium nuclei. 
The confidence level, however, is too low to claim discovery 
and the events might turn out to be misidentified background.
Nonetheless, if these results
are confirmed by other ongoing and in the future by more sensitive 
direct detection experiments, it would be a huge boost for the 
lightest supersymmetric particle (LSP)
of the minimal supersymmetric standard model (MSSM), 
which is by far the leading and most compelling DM candidate.

In this paper, partly inspired by the CDMS II experiment, we
investigate neutralino LSP dark matter in the context of flipped
SU(5)~\cite{DeRujula:1980qc}, a grand unified theory (GUT) which is closely related to
but has some important differences from SU(5). One
important advantage flipped SU(5) (FSU(5)) has over other GUTs 
such as SU(5) and SO(10) is the
remarkable ease with which the doublet-triplet splitting problem
is resolved~\cite{ant} within a minimal Higgs framework. 
Another crucial advantage is the simplicity 
with which successful minimal supersymmetric hybrid inflation can be
implemented in FSU(5) \cite{Rehman:2009yj}. Motivated in part by 
these an other important features
we have explored the Higgs and sparticle spectroscopy of MSSM by
embedding it within the FSU(5) framework supplemented by $N=1$
supergravity.

We focus in particular on finding a relatively light 
sparticle spectrum and neutralino dark matter which is compatible
with the recent CDMS II results. For some other recent attempts 
in the framework of supersymmetry see \cite{Feldman:2009pn}. 
To implement our strategy the magnitudes of the 
soft supersymmetry breaking mass parameters are
kept below a TeV, and comparison is made with the constrained 
minimal supersymmetric standard model (CMSSM) \cite{Feldman:2008hs} and non-universal 
Higgs mass (NUHM2) \cite{Baer:2008ih} models after
similar restrictions on their soft mass parameters are imposed. With FSU(5) enabled to 
capture suitable features from both CMSSM and
NUHM2, attractive neutralino dark matter candidates with mass 
$\gtrsim 30{\rm GeV}$ and fully consistent with the CDMS II bounds are
obtained. 

The suppersymmetric FSU(5) model is based on the maximal
subgroup $G\equiv {\rm SU(5)}\times {\rm U(1)}_X$ of SO(10), and the sixteen
chiral superfields per family of SO(10) are arranged under $G$ as: 
$10_1 =  ( d^c,  Q, \nu^c)$, $\bar{5}_{-3} = (u^c , L)$, 
$1_5= e^c$. Here the subscripts refer to the respective charges under 
${\rm U(1)}_X$, and we follow the usual notation for 
the Standard Model (SM) particle
content. The MSSM electroweak Higgs doublets $H_u$ and $H_d$ belong
to $\bar 5_H$ and $5_H$ of SU(5), respectively. We will assume for
simplicity that the soft mass$^2$ terms, induced at $M_{\rm GUT}$ through
gravity mediated supersymmetry breaking \cite{Chamseddine:1982jx},
are equal in magnitude for the scalar squarks and sleptons of the
three families. The asymptotic MSSM gaugino masses, on the other
hand, can be non-universal. Due to the FSU(5) gauge structure
the asymptotic ${\rm SU(3)}_c$ and ${\rm SU(2)}_W$ gaugino masses can be
different from the ${\rm U(1)}_Y$ gaugino mass. Assuming SO(10) normalization for
${\rm U(1)}_X$, the hypercharge generator in FSU(5) is given
by $Y=(-{Y_5}/2+ \sqrt{24} X)/5$, where $Y_5$
and $X$ are the  generators of SU(5) and ${\rm U(1)}_X$ \cite{Barr:2006im}. We 
then have the following asymptotic relation between the three MSSM gaugino masses:
\begin{align}
M_1=\frac{1}{25} M_5 + \frac{24}{25} M^\prime,\; \mbox{with} \;
M_5=M_2=M_3, \label{gauginoCondition}
\end{align}
where $M_5$ and $M^\prime$ denote SU(5) and ${\rm U(1)}_X$ gaugino
masses respectively. The supersymmetric FSU(5) model  thus has two
independent parameters $(M_2=M_3,\, M^\prime)$ in the gaugino
sector. In other words, in FSU(5), by assuming gaugino
non-universality, we increase by one the number of fundamental
parameters compared to the CMSSM. 

We will also consider both universal (${m^2_{H_u}}$$=$${m^2_{H_d}}$) 
and non-universal (${m^2_{H_u}}$$\neq$${m^2_{H_d}}$) soft scalar
Higgs masses in FSU(5), which would mean up to three additional parameters compared to
the CMSSM. This latter case, with one additional gaugino mass parameter and two soft
scalar mass parameters, provides us with some of the most compelling neutralino
dark matter candidates for direct detection in the ongoing and future
experiments. 

We use the ($\mu,m_A$) parametrization to characterize non-universal 
soft scalar Higgs masses rather than ($H_u,H_d$), where $\mu$ denotes the 
coefficient of the supersymmetric bilinear term involving $m_{H_u}$ and 
$m_{H_d}$, and $m_A$ is the mass of the CP odd scalar boson. The fundamental parameters of our 
FSU(5) model are;
\begin{align}
m_0,M^\prime,M_2,\tan\beta,A_0,\mu,m_A, \label{params}
\end{align}
where $m_0$ is the scalar soft mass, $M^\prime$ and $M_2$ are the soft gaugino 
masses discussed above, $\tan\beta$ is the ratio of the 
vacuum expectation values (VEVs) of the two 
Higgs doublets, $A_0$ is the universal trilinear soft term, and 
we will assume that ${\rm sgn}\,\mu>0$. Note that $\mu$ and
$m_A$ are specified at the weak scale, whereas the other parameters are
specified at $M_{\rm GUT}$. Although not required, we will assume that 
the gauge coupling unification condition $g_3=g_1=g_2$ holds at $M_{\rm GUT}$ 
in FSU(5). 


We use the ISAJET~7.79 package~\cite{ISAJET} to perform random scans over
the FSU(5) parameter space shown in Eq. (\ref{params}). ISAJET employs two-loop MSSM renormalization
group equations (RGEs) and defines $M_{\rm GUT}$ to be the scale at which $g_1=g_2$.
This is more than adequate as a few percent deviation
from the exact unification condition $g_3=g_1=g_2$ can be assigned to unknown
GUT-scale threshold corrections~\cite{Hisano:1992jj}.

We have performed random scans for the following parameter range:
\begin{align}0\leq  m_{0} \leq 1\, \rm{TeV}, \nonumber \\
0\leq  M^\prime \leq 1\, \rm{TeV}, \nonumber  \\
0\leq  M_{2} \leq 1\, \rm{TeV}, \nonumber  \\
0 \leq m_{A} \leq 1\, \rm{TeV}, \nonumber \\ 
0 \leq \mu  \leq 10\, \rm{TeV}, \nonumber \\
\tan\beta=10,30,50 \nonumber \\ 
A_{0}=0,
\label{parameterRange}
\end{align}
with $m_t = 173.1\, {\rm GeV}$ \cite{:2009ec}. The results are not too 
sensitive to one or two sigma variation in the value of $m_t$.

We also collected data for the CMSSM and NUHM2 models as well as for
FSU(5) with universal Higgs boundary conditions 
(FSU(5)-UH) in order to study the impact on the CMSSM predictions if gaugino 
non-universality subject to Eq. (\ref{gauginoCondition}) is imposed.

\begin{figure}[t]
\centering
\includegraphics[width=8.5cm]{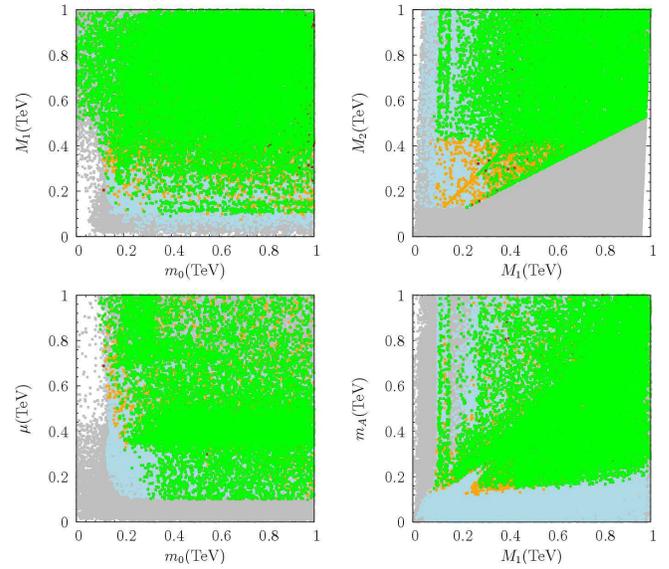}
\caption{
Plots in the $M_1$ - $m_0$, $M_2$ - $M_1$, $\mu$ - $m_0$ and $m_A$ - $M_1$
planes for the FSU(5)
model for $\tan\beta=30$. Gray points are consistent with REWSB and $\tilde{\chi}^0_{1}$ LSP.
Light blue points satisfy the WMAP bounds on $\tilde{\chi}^0_1$ dark matter abundance
and particle mass bounds, except the bound on the lightest Higgs mass.
Orange points form a subset of light blue points that satisfies
constraints from $BR(B_s\rightarrow \mu^+ \mu^-)$ and $BR(b\rightarrow s \gamma)$.
Brown points belong to the subset of orange points that satisfies the
LEP2 bound on the lightest Higgs mass. Green points form a subset of brown points
that satisfies all constraints including $\Delta (g-2)_\mu$.
\label{fund}}
\end{figure}

While scanning the parameter space, we employ the Metropolis-Hastings
algorithm as described in \cite{Belanger:2009ti}. All of the collected data points satisfy 
the requirement of radiative electroweak symmetry breaking (REWSB) 
with the neutralino in each case being the LSP. Furthermore, 
all of these points satisfy the constraint $\Omega_{\rm CDM}h^2 \le 10$.
This is done so as to collect more points with a WMAP compatible value of cold dark
matter (CDM) relic abundance. For the Metropolis-Hastings algorithm, we only use
the value of $\Omega_{\rm CDM}h^2$ to bias our search. Our purpose in using the
Metropolis-Hastings algorithm is to be able to search around regions of
acceptable $\Omega_{\rm CDM}h^2$ more fully. After collecting the data, we impose 
the mass bounds on all the particles~\cite{Amsler:2008zzb} and use the
IsaTools package~\cite{Baer:2002fv}
to implement the following phenomenological constraints:
\begin{table}[h]\centering
\begin{tabular}{rlc}
$m_h~{\rm (lightest~Higgs~mass)} $&$ \geq\, 114.4~{\rm GeV}$                    &  \cite{Schael:2006cr} \\
$BR(B_s \rightarrow \mu^+ \mu^-) $&$ <\, 5.8 \times 10^{-8}$                     &   \cite{:2007kv}      \\
$2.85 \times 10^{-4} \leq BR(b \rightarrow s \gamma) $&$ \leq\, 4.24 \times 10^{-4} \; (2\sigma)$ &   \cite{Barberio:2007cr}  \\
$\Omega_{\rm CDM}h^2 $&$ =\, 0.111^{+0.028}_{-0.037} \;(5\sigma)$               &  \cite{Komatsu:2008hk}    \\
$3.4 \times 10^{-10}\leq \Delta (g-2)_{\mu}/2 $&$ \leq\, 55.6 \times 10^{-10}~ \; (3\sigma)$ &  \cite{Bennett:2006fi}
\end{tabular}

\end{table}


In Fig.~\ref{fund} we present the results for the FSU(5) model 
in the $M_1$ - $m_0$, $M_2$ - $M_1$, $\mu$ - $m_0$ and $m_A$ - $M_1$ planes 
for $\tan\beta=30$. Points consistent with REWSB and $\tilde{\chi}^0_{1}$ dark matter 
with $\Omega_{\rm CDM}h^2 \leq 10$ are shown in gray. Light blue points satisfy the 
WMAP bounds on $\tilde{\chi}^0_1$ dark matter abundance
and particle mass bounds except the bound on the lightest Higgs mass. A subset of light blue points, shown in 
orange, satisfies constraints from $BR(B_s\rightarrow \mu^+ \mu^-)$ and $BR(b\rightarrow s \gamma)$. 
Shown in brown is the subset of orange points that satisfies the LEP2 bound on the 
lightest Higgs mass. Finally, we show points in green that are a subset of the brown points 
and which satisfy all experimental constraints including $\Delta (g-2)_\mu$. The $M_1$ - $m_0$ plane is in sharp 
contrast with the corresponding plane for the CMSSM with fixed $\tan\beta$ and $A_0$. 
\begin{figure}[b]
\centering
\includegraphics[width=8.5cm]{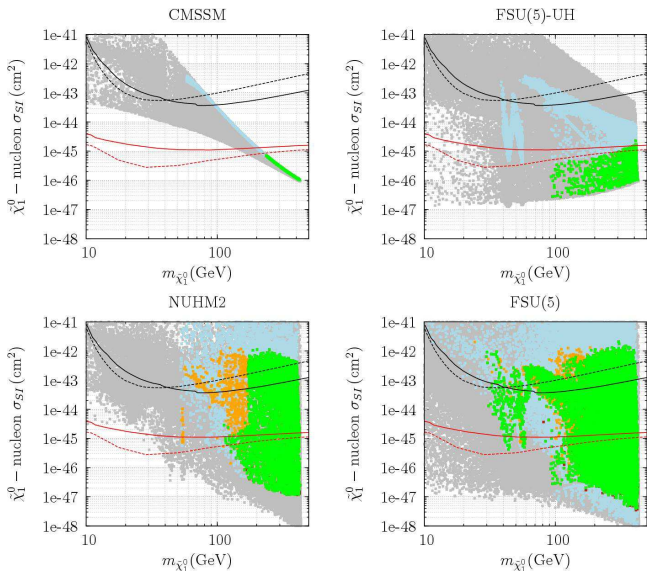}
\caption{
Plots of the $\tilde{\chi}^0_1$-nucleon spin-independent cross-section versus
$m_{\tilde{\chi}^0_1}$ for the CMSSM, NUGM, NUHM2 and FSU(5)models for $\tan\beta=30$.
Color coding same as in Fig.~\ref{fund}. Also shown are current limits from
CDMS II (solid black line), XENON10 (dashed black) and projected reach of 
SuperCDMS (solid red) and XENON100 (dashed red). Note that both axes are logarithmic. 
\label{tb30sigSI}}
\end{figure}
Non-universality in the Higgs sector opens up the $A$ funnel region where resonant annihilation 
occurs via the CP odd Higgs $A$. We also have the $Z$ pole and lightest Higgs resonance 
annihilation regions. Gaugino nonuniversality, subject to Eq. (\ref{gauginoCondition}), allows for 
the bino-wino coannihilation channel \cite{Baer:2005jq}. Bino-wino coannihilation occurs for $2 M_2 \sim M_1$ at $M_{\rm GUT}$ 
since at 1-loop $M_1$ renormalizes by a factor $\sim 2$ larger than $M_2$. This possibility can be seen in 
the $M_2$ - $M_1$ plane in Fig.~\ref{fund}. The line $M_2=M_1$ corresponds to the NUHM2 model, 
and off-diagonal allowed regions are a manifestation of non-universality in the gaugino sector. The 
region $M_2\lesssim M_1/2$ is disfavored because of the lower bound on dark matter relic abundance and 
in this region the wino, which has a large annihilation cross section, becomes 
a significant component of the neutralino. The white region in the upper left corner of this 
$M_2$ - $M_1$ plane occurs because we only allow $M^\prime > 0$, so that an artificial lower 
bound of $M_1=M_2/25$ is imposed. 

The $\mu$ - $m_0$ plane of 
Fig.~\ref{fund} shows the very interesting possibility of a very low value of 
$\mu \sim 100\, {\rm GeV}$ for $m_0\gtrsim 300\, {\rm GeV}$. In the $m_A$ - $M_1$ plane one can see the 
somewhat sharp bound $m_A$$\sim$$180\, {\rm GeV}$. This, as well as the lower bound on $M_1$, is 
caused primarily by a contribution to $BR(B_s\rightarrow \mu^+ \mu^-)$ or 
$BR(b\rightarrow s \gamma)$ which is too high 
because of a relatively light spectrum.

\begin{table}[h!]
\centering
\scalebox{0.8}{
\begin{tabular}{lccccc}
\hline
\hline
                 & Point 1 & Point 2 & Point 3 & Point 4 & Point 5      \\
\hline
$m_{0}$          &   399      & 704   &  247   & 496   &  937     \\
$M_{1} $         &   110      & 147   &  186   & 312   &  373  \\
$M_{2} $         &   893      & 708   &  484   & 714   & 780  \\
$\tan\beta$      &   30       &  30   & 30     & 10    & 50     \\
$A_0$            &    0     &  0  &  0   &  0 & 0    \\
$\mu$            &    101     & 184   &  662   & 153 & 293     \\
$m_{A} $         &    451     &  351  &  183   & 597 & 433     \\

\hline
$m_h$            &  119    & 117     & 115    &  116 & 118     \\
$m_H$            &  454    & 354     & 184     & 601 & 435      \\
$m_{H^{\pm}}$    &  462    & 363     & 202     & 606 & 445     \\

\hline
$m_{\tilde{\chi}^0_{1,2}}$
                 & 31,112  & 53,186     & 73,378   & 105,162 & 150,295     \\
$m_{\tilde{\chi}^0_{3,4}}$
                 & 113,726 & 194,582    & 665,675   & 172,586 & 303,646  \\

$m_{\tilde{\chi}^{\pm}_{1,2}}$
                 &105,717  & 187,574  & 379,676  & 155,578 & 298,637 \\
$m_{\tilde{g}}$  & 1960    &  1600    & 1110 &  1600   & 1760  \\

\hline $m_{ \tilde{u}_{L,R}}$
                 & 1840,1770  & 1610,1570  & 1050,1020  & 1540,1500 & 1840,1790   \\
$m_{\tilde{t}_{1,2}}$
                 & 1300,1630  & 1110,1410  & 814,1000   & 1070,1390 & 1260,1510  \\
\hline $m_{ \tilde{d}_{L,R}}$
                 & 1840,1750 & 1610,1550 & 1050,1010  & 1540,1480 & 1840,1780  \\
$m_{\tilde{b}_{1,2}}$
                 & 1600,1670 & 1380,1470  & 943,995  & 1370,1470 & 1480,1510  \\
\hline
$m_{\tilde{\nu}_{1}}$
                 & 719       & 853      &  406      &  695  & 1060  \\
$m_{\tilde{\nu}_{3}}$
                 & 698       & 829      &  410      &  690     & 929  \\
\hline
$m_{ \tilde{e}_{L,R}}$
                & 726,292    & 858,658 & 415,218  & 701,447 & 1070,933  \\
$m_{\tilde{\tau}_{1,2}}$
                & 204,706    & 599,834  & 219,431  & 444,698  & 574,933   \\
\hline

$\sigma_{SI}({\rm cm}^2)$
                & 5.19$\times 10^{-44}$ & 4.64$\times 10^{-44}$ & 3.51$\times 10^{-44}$
				& 3.23$\times 10^{-44}$ & 4.13$\times 10^{-44}$   \\

$\Omega_{CDM}h^2$
                & 0.115      & 0.098    &  0.119     &  0.118   & 0.098 \\
\hline
\hline
\end{tabular}
}
\caption{ Sparticle and Higgs masses (in GeV units),
with $m_t=173.1\, {\rm GeV}$. These five points, with $\tan\beta$ from $10$ to $50$, 
satisfy all the constraints and have a $\sigma_{SI}$ that is slightly below the CDMS II bound in each case. 
\label{table1}}
\end{table}

In Fig.~\ref{tb30sigSI} we present plots of $\tilde{\chi}^0_1$-nucleon
spin-independent cross-section ($\sigma_{SI}$) versus $m_{\tilde{\chi}^0_1}$ for the
CMSSM, FSU(5)-UH, NUHM2 and FSU(5) models for $\tan\beta=30$. It is clear that
for the choice of parameters in Eq. (\ref{parameterRange}), only the FSU(5) model
can provide a suitable dark matter candidate at or near the upper bound on
$\sigma_{SI}$ set by CDMS II (solid black line in Fig.~\ref{tb30sigSI}).
The CMSSM can 
provide a suitable $\tilde{\chi}^0_1$, and it does so in the focus point region
where $m_0$ is large ($ > 1\, {\rm TeV}$), and $\tilde{\chi}^0_1$ acquires
a significant higgsino component because of small $\mu$. The FSU(5)-UH model,
with $M_1$ as a free parameter, does slightly better than
the CMSSM by allowing lighter neutralinos to achieve a relatively
higher $\sigma_{SI}$. However, both the CMSSM and FSU(5)-UH models suffer from
a $BR(b\rightarrow s \gamma)$ which is too large. This problem is cured in the
NUHM2 and FSU(5) models by allowing $\mu$ to be a free parameter, thereby
allowing for the possibility to suppress or cancel the contributions
to $BR(b\rightarrow s \gamma)$. The NUHM2 model, however, is disfavored for 
our choice of parameters and for $m_{\tilde{\chi}^0_1} \lesssim 130 \,{\rm GeV}$ 
because of the LEP2 bound on the mass of the lightest Higgs boson. 
The FSU(5) model cures this problem by decoupling $M_3$ from $M_1$ and, therefore, allowing for
a heavier gluino than in the NUHM2 case for the same neutralino mass. The
contributions to the squark masses from the gluino loops are responsible for
heavier squark masses in the FSU(5) model which, in turn, provide for a relatively
heavier Higgs mass. We can see in Fig.~\ref{tb30sigSI} that, from the point of view of CDMS II, 
FSU(5) provides a satisfactory dark matter candidate, consistent with all the constraints for 
$m_{\tilde{\chi}^0_1}\gtrsim 30\, {\rm GeV}$.
The results for $\tan\beta=10,50$ are similar. Also shown in Fig.~\ref{tb30sigSI}
are the current bounds on $\sigma_{SI}$ from XENON10 (dashed black line)
and the projected reach of SuperCDMS and XENON100
(solid and dashed red lines, respectively).

In Table~\ref{table1} we present five benchmark points ranging in mass from
$30\, {\rm GeV}$ to $150\, {\rm GeV}$ and in each case the cross section
is just below the bound set by CDMS II. Points 1 through 3 correspond
to $\tan\beta=30$, point 4 has $\tan\beta=10$ and point 5 has $\tan\beta=50$.
These points satisfy all the collider and astrophysical constraints. 
If the CDMS II results are confirmed, Table~\ref{table1} presents 
predictions for sparticle
and Higgs spectroscopy which can be tested at the LHC.


Motivated by the newly released CDMS II upper bounds on spin independent 
WIMP-nucleon cross sections, and with new experiments of even greater 
sensitivity expected to go online in the near future, we have examined 
the soft supersymmetry breaking parameter space of MSSM embedded in 
flipped SU(5) to identify realistic neutralino dark matter candidates. We 
have identified regions of the parameter space containing perfectly 
viable neutralino dark matter candidates ranging in mass from around $30$ 
to more than $100\, {\rm GeV}$. The corresponding sparticle and Higgs mass spectra 
is also presented with several new particles likely to be accessible at 
the LHC.

We thank S.M. Barr and Yukihiro Mimura for valuable discussions. This work 
is supported in part by the DOE Grant No. DE-FG02-91ER40626 
(I.G., R.K., S.R. and Q.S.) and GNSF Grant No. 07\_462\_4-270 (I.G.).

\end{document}